# Grain Boundary Loops in Graphene

Eric Cockayne,[1][§] Gregory M. Rutter,[2,3] Nathan P. Guisinger,[3][*] Jason N. Crain,[3]

Phillip N. First,[2] and Joseph A. Stroscio[3][§]

[1]Ceramics Division, NIST, Gaithersburg, MD 20899
[2]School of Physics, Georgia Institute of Technology, Atlanta, GA 30332
[3]Center for Nanoscale Science and Technology, NIST, Gaithersburg, MD 20899

Topological defects can affect the physical properties of graphene in unexpected ways. Harnessing their influence may lead to enhanced control of both material strength and electrical properties. Here we present a new class of topological defects in graphene composed of a rotating sequence of dislocations that close on themselves, forming grain boundary loops that either conserve the number of atoms in the hexagonal lattice or accommodate vacancy/interstitial reconstruction, while leaving no unsatisfied bonds. One grain boundary loop is observed as a "flower" pattern in scanning tunneling microscopy (STM) studies of epitaxial graphene grown on SiC(0001). We show that the flower defect has the lowest energy per dislocation core of any known topological defect in graphene, providing a natural explanation for its growth via the coalescence of mobile dislocations.

---

[§]Authors to whom correspondence should be addressed:eric.cockayne@nist.gov, joseph.stroscio@nist.gov

[*]Present address: Argonne National Laboratory, Argonne, IL

I. **Introduction**

The symmetry of the graphene honeycomb lattice is a key element for determining many of graphene's unique electronic properties. The sub-lattice symmetry of graphene gives rise to its low energy electronic structure, which is characterized by linear energy-momentum dispersion.[1] The spinor-like eigenstates of graphene lead to one of its celebrated properties, reduced backscattering (*i.e.*, high carrier mobility), which results from pseudo-spin conservation in scattering within a smooth disorder potential.[2,3] Topological lattice defects break the sublattice symmetry, allowing insight into the fundamental quantum properties of graphene. For example, defects allow a detailed study of how symmetry breaking relates to pseudo-spin (non)conservation, as shown in scanning tunneling microscopy studies of inter- and intra-valley scattering in graphene in the presence of local defects.[4-7] Defects have profound effects in the chemical, mechanical, and electronic properties of graphene.[8,9] Atomistic calculations indicate that graphene with large angle tilt-grain boundaries composed of an array of defects is as strong as pristine material, and much stronger than graphene with low-angle boundaries.[10] Recent measurements and analysis suggest that some of these one-dimensional extended defects can have unique electronic properties, such as a one-dimensional conductivity.[11]

While point defects, such as vacancies[12] and single impurity atoms[13] have been studied in the graphene lattice, extended topological defects have only recently been investigated.[10,14-17] Extended topological defects in the hexagonal lattice can form from networks of five- and seven-membered rings, which, individually, are disclinations in the hexagonal lattice that preserve 3-fold bonding.[15,18] The energetics of extended topological defects are of interest in creating two-dimensional defect patterns,[19] which may be useful in chemical sensing or mechanical and electronic applications.

In this article we show that nonlinear topological defects are formed by incorporating five- and seven-membered rings into the pristine graphene lattice in closed loops, resulting in closed grain boundary loops. We outline the construction of the grain boundary loops in terms of the primitive five- and seven-membered rings. Our focus is on closed grain boundary loops that are equivalent to cutting out a portion of the hexagonal lattice and rotating it relative to the pristine graphene lattice, and hence we also refer to these defects as *rotational* grain boundaries. Density functional theory (DFT) calculations are used to examine the energetics of rotational grain boundaries. Remarkably, the construction consisting of close-packed five- and seven-membered rings with hexagonal symmetry (the "flower" defect described below) is predicted to have the lowest energy per dislocation core of any known topological defect in graphene. Therefore the formation of rotational grain boundaries might be favorable under certain growth conditions. Previous STM measurements of epitaxial graphene on SiC(0001)[4,20] observed a unique defect with 6-fold symmetry, which was not identified. Our present study reveals that this structure is a low-energy rotational grain boundary formed in the high temperature growth, identical to the flower defect discovered by transmission electron microscopy on CVD graphene.[21,22] Simulated STM topographs of this rotational grain boundary are in excellent agreement with experimental STM images. Scanning tunneling spectroscopy (STS) measurements show that this defect has an electronic state at 0.5 eV above the charge neutrality point (Dirac point) and DFT calculations of the density of electron states show a localized resonance in the same energy range.

## II. Experimental and Computational Methods

The epitaxial graphene samples studied here were grown on SiC wafers in ultra-high vacuum (UHV), which typically produces more defects than growth by the induction furnace method.[23] Two groups of SiC(0001) (silicon face) epitaxial graphene samples were measured in two custom STM systems at NIST. The first group of samples was grown at Georgia Tech on $H_2$-etched surfaces of high-purity semi-insulating 4H-SiC by thermal desorption of silicon in UHV. After transport in atmosphere and transfer of the samples into the UHV system at NIST, the samples were out-gassed at 800 °C and STM measurements made at 4.2 K. This treatment produced a number of defects with different symmetries,[4] including the flower defect discussed in this article. A second set of samples was grown on 6H-SiC(0001) by resistively heating as-received SiC wafers in UHV, followed by STM measurements at 295 K without exposure to atmosphere. The second set of samples contained far more flower defects, including regions with networks of them.[20] Further details of these samples are described in previous publications.[4,20]

The computational studies were performed using the density functional theory code Vienna Ab-initio Simulation Package (VASP).[24] Ultrasoft pseudopotentials[25] for the carbon core electrons, together with the local density approximation for the exchange-correlation functional, were used throughout. A rather small plane wave cutoff of 211 eV was used due to large supercell sizes; studies with larger cutoffs showed that the features of the electronic density become sharper, but that there is no significant change in the calculated density of states (DOS). Brillouin zone integration for relaxation was effected using k-point grids with approximately $5000/N_C$ k-points in the full Brillouin zone ($N_C$ the number of carbons in the calculation); for DOS calculations, approximately $200000/N_C$ k-points were used. Various periodic graphene monolayer and multilayer supercells containing one rotational defect were investigated. In each case, the supercell height was set to 1.34 nm to avoid spurious interactions between image layers. Cell parameters were relaxed such that the in-plane Pulay stress was equal to the stress of graphite at the experimental lattice parameters under the same computational parameters. After relaxation of the atomic structure, electronic densities of states calculations were performed. Images of the projected density of states as a function of spatial position are used to simulate STS d$I$/d$V$ maps. STM topographic images were simulated using the Tersoff and Hamann approximation[26] which relates the STM image to the integrated electronic density of states between the Fermi level and the bias voltage.

### III. Grain Boundary Defects

#### A. Linear grain boundaries

A linear grain boundary in graphene consists of a chain of dislocations, each of which can be decomposed into a pair of five- and seven-membered carbon rings.[15] The topological defect comprising a ring of five atoms at the defect core (see Fig. 1(a)) is equivalent to subtracting an infinite 60 degree wedge of graphene from the unperturbed hexagonal lattice (and is also called a +60 degree disclination). Every carbon atom is bonded to three others and so there are no unsatisfied bonds. Similarly, in Fig. 1(b) we show a –60 degree disclination which has a seven-atom ring at its core. By combining these two types of topological defects, we can generate more interesting and extended cases. For example, in Fig. 1(c) we show a dislocation constructed from a five-atom ring sharing atoms with a seven-atom ring, at the termination of a semi-infinite strip of atoms inserted into the lattice. The dislocation is described by a Burgers vector, **b**($p,q$), where the components ($p,q$) can be identified from the lattice translation vectors.[15]

Aligning a chain of (0,1)-dislocations gives a linear grain boundary. Distinct dense packing of these (0,1)-dislocations creates a large angle grain boundary with angle $\theta=21.8°$ (Fig. 1(d)), while packing of (0,1) + (1,0) dislocations creates a large angle grain boundary with angle $\theta=32.2°$ (Fig. 1(e)). These defects were recently identified as low-energy structures[15] and as ones that can enhance the mechanical strength of graphene compared to low angle boundaries with fewer defects.[10]

### B. Grain boundary loops

If sequential dislocations are rotated (Fig. 1(f)) and arranged in a closed loop, a grain boundary loop is formed (Fig. 2). Different grain boundary loops differ in the magnitude of the disclinations involved (*i.e.* the number of carbon atoms in the rings), and the geometry of their arrangements around the loop. An important subset of grain boundary loops, the rotational grain boundaries, is generated by cutting out a portion of the graphene lattice, rotating it, and then "gluing" it back in. This procedure naturally conserves the number of carbon atoms with respect to ideal graphene.

We have identified several low-energy rotational grain boundaries in graphene (Fig. 2), as well as one family of rotational grain boundaries with six-fold symmetry (Fig. 3). The Stone-Wales defect (Fig. 2(a)) is the smallest rotational grain boundary in graphene and has a $C_2$ rotational axis. A $C_3$ symmetric structure results if we combine three dislocations with successive Burgers vectors rotated by 120° relative to one another (Fig. 2(b)). Finally, if we combine six dislocations with successive Burgers vectors rotated by 60° increments, then the two dimensional defect structure closes on itself with a $C_6$ symmetry (Fig. 2(c)). Other topological defects in graphene with closed loops of dislocation cores have been reported in the literature[27-33] but do not conserve the number of C atoms (Fig. 3). Such grain boundary loops can be viewed as a combination of a rotational grain boundary with a vacancy/interstitial reconstruction. In Fig. 3(a), the defect removes two atoms, while in Fig. 3(b), the defect adds two atoms. To a first approximation, grain boundary loops that conserve or reduce the number of atoms are expected to remain flat, while those that increase the number of atoms are expected to buckle out of the plane to accommodate the stress of the locally high atom density. Such out-of-plane warping is indeed found for the defect in Fig. 3(b).[27]

The rotational grain boundary in Fig. 2(c) is the smallest member of the family of "$C_6(m,n)$" defects, with $m = n = 1$. Other members of this family are shown in Fig. 4. We label the smallest member of this family the "flower" defect based on STM observations[4,20] and on our demonstration in this work of its equivalence to the flower defect observed in CVD grown graphene.[21,22] In this family, the central $6(m+n)^2$ carbon atoms are rotated by an angle of approximately $(n/(m+n))$ 60° (exactly 30° for $m = n$). The flower defect has the lowest energy of the rotational grain boundaries shown in Fig. 2 with the exception of the Stone-Wales defect in Fig. 2(a). C-C bond lengths range from 138.5 pm to 144.3 pm, as compared to the ideal graphene value of 142 pm. The C-C-C angles range from 104.5° to 135.9° compared with 120° in ideal graphene, 108° in a regular pentagon, and 128.6° in a regular heptagon. The flower rotational grain boundary is stable with respect to out-of-plane distortions and is calculated to have an energy of +7.0 eV above ideal graphene; higher than the +4 eV calculated energy of a Stone-Wales defect,[27,34] but smaller than the +10 eV energy of an (unreconstructed) carbon vacancy (see Table 1). At 1.2 eV per dislocation (*i.e.* per 5-7 pair), the flower defect has a lower energy per dislocation than the Stone-Wales defect (2.0 eV per 5-7 pair), and indeed, lower than the energy of any other known topological defect in graphene (see Table 1). This suggests a natural explanation for the growth of the flower defect (and other low energy grain boundary loops) via the coalescence of mobile dislocations or Stone-Wales defects. The $C_3$ defect shown in Figure 2(b) has a higher energy, but might be obtained under certain growth conditions. The similarity in energy per dislocation core between the flower defect and the large angle grain boundary in Fig. 1(e) suggests that arrangements of dislocation cores in which all five-membered rings share bonds with two seven-membered rings, and vice versa, may be particularly stable.

Each rotational grain boundary studied expands the lattice. The flower defect expands the lattice by 0.022 nm$^2$ with respect to an ideal graphene lattice with the same number of atoms, and has the smallest ratio of energy change to expansion area of all rotational grain boundaries studied. Thus, it would be interesting to investigate the possible role of this defect in graphene under tensile strain, and in graphene fracture.[10]

### C. STM observations and density of states of the flower defect

Defects have been observed in STM imaging of graphene and graphite.[4,35] Typically, 3-fold patterns are observed on the surfaces of graphite, and are understood in terms of scattering involving defects in one of the two 3-fold sublattices that make up the honeycomb lattice.[35,36] A unique defect with 6-fold symmetry is observed for graphene grown epitaxially on SiC (Figs. 5(a) and 5(c)), but its origin has not yet been determined.[4,7,20,37] In some cases this defect is observed to cluster in groups,[20] and it remains an open question to account for the appearance of these defects during graphene growth. To the best of our knowledge this 6-fold pattern has not been observed in STM images of graphite surfaces. It cannot be explained by a simple lattice defect in one of the graphene sublattices, as this would give rise to a 3-fold pattern in STM images, nor by a Stone-Wales defect, which gives rise to 2-fold symmetry in STM topographs.[38]

DFT simulations of the flower $C_6(1,1)$ rotational grain boundary, as shown in Fig. 5(b), bear a striking resemblance to the 6-fold pattern observed in the STM image of Figs. 5(a) and 5(c). The simulation reproduces all essential experimental features such as: a dark center, a nearly circular ring of intensity in the central region, and the 6-fold symmetric dark spokes radiating from the center. An examination of the overlaid lattice structure shows the intensity of the inner central region arises from the radial C-C bonds fanning out of the center hexagon. The second circular high intensity region arises from the C-C bonds pointing radially outward from the center that are shared by the pentagons and heptagons. The $\sqrt{3} \times \sqrt{3}$ R30° intensity modulation outside the defect arises from the scattering between low-lying K and K' states in graphene.[4] While the creation of the $\sqrt{3} \times \sqrt{3}$ R30° modulated features in the electronic structure of graphene due to defects is well known, the quantum phase factor of the coupling is sensitive to the nature of the defect; the presence of circular rings over several tenths of eV seems to be characteristic of the flower defect both experimentally and computationally. From the above observations, we conclude that the previously unidentified STM flower defect in graphene is the rotational grain boundary shown in Fig. 2(d). Experimentally observed symmetry breaking from six-fold to three-fold, visible especially in the center of Fig. 5(a), arises from the weak interaction with a second, subsurface, graphene layer in the experiment; this effect is reproduced in simulations of a flower defect in a Bernal-stacked bilayer (not shown).

Topological defects and grain boundaries in graphene are well known to produce electronic states close to the charge neutrality (Dirac) point.[12,15,39,40] Scanning tunneling spectroscopy measurements on the flower defect show an electronic state at 0.5 eV above the Dirac point localized to the region of the defect (Figs. 6(a) and 6(b)). A state at similar energies is observed in the large-angle linear grain boundaries,[15] and in the Stone-Wales defect.[28] We would not expect this state to be the same in the flower defect due to symmetry considerations.

However, the calculated DOS for this defect (Fig. 6(d)) shows that related peaks do occur at similar energies. In particular, two peaks are seen at 0.2 eV and 0.4 eV above the Dirac point, close to the experimental peak. The appearance of two peaks in this energy range is an artifact of the finite size of the simulation, and there is, in fact, only one underlying resonance at about $E_D + 0.3$ eV, as discussed below. This resonance corresponds to states with intensity localized on the flower defect, as seen in the spatially projected density of states in Fig. 6(c), and is largely responsible for the intensity modulation observed in the STM simulation of Fig. 5(b). The experimental spectrum in Fig. 6(b) shows only a single broad feature at 0.5 eV above the Dirac point, which corresponds well to the calculated resonance at $E_D + 0.3$ eV.

A tight-binding model with bond-length-dependent couplings[41] was fit to the DFT results and then used to investigate the effect of supercell size on the peak structures at $E_D + 0.3$ eV. Larger supercells were created by padding the DFT-relaxed structure of a smaller supercell with ideal graphene. The results show quite significant finite size effects even at the size of the largest DFT cell (864 atoms). In particular, a set of resonances of fixed energy (with respect to supercell size) is observed inside the flower defect. These resonances then couple to those external wavefunctions that have significant projection onto the boundary of the flower defect. These external wavefunctions have energies that vary as $r^{-1}$ ($r$ the linear dimension of the supercell), and couple less strongly to the flower defect as $r$ increases. The tight-binding model shows that there are actually three resonances near $E_D$, not four, at energies of $E_D - 0.4$ eV (strongest; $E_{2g}$ symmetry of point group $D_{6h}$), $E_D + 0.3$ eV (next strongest; $E_{2g}$), and $E_D + 0.6$ eV (weak; $B_{1u}$). Coincidentally, the 864-atom cell has an external state very close in energy to $E_D + 0.3$ eV that couples within the internal resonance and splits it into two peaks. Further support for this conclusion, based entirely on the DFT calculations rather than tight-binding models, is given by (1) the similarity of the wavefunctions corresponding to peaks 2 and 3 (Fig. 7), and (2) the fact that the states that yield peaks 2 and 3 do not extend throughout the Brillouin zone, but come from complementary regions. The predicted weakness of the resonance at $E_D + 0.6$ eV, along with the resolution limit of the experiment due to thermal broadening, probably explains why this higher-energy peak is not observed experimentally.

Figure 7 shows the spatial projected density of states of the four main peaks in the DOS near $E_D$, as could be measured by STS d$I$/d$V$ mapping. Interestingly, the predicted spatial mapping of these resonant states (Figure 7) show a variety of different nodal structures, which can be used to distinguish the various states based on symmetry and intensity distributions. In particular, the map corresponding to the state at $E_D$-0.4 eV has a dramatically different structure than the others, with strong intensity along the pentagon ring axes. This energy is outside the range of the measurements in Fig. 6(b), and awaits future experimental investigations.

In summary, we have described a type of topological defect in graphene which consists of a closed loop of disclinations (grain boundary loops), and an important subset of grain boundary loops (rotational grain boundaries) that corresponds to a rotation of the honeycomb lattice within the core of the defect with respect to the surrounding lattice, preserving three-fold coordination of the carbon atoms. A theoretical study of the energies of rotational grain boundaries, originating from paired five- and seven-membered ring disclinations, shows that the previously unidentified "flower" defect in STM topographs is the smallest member of a family of rotational grain boundary defects with $C_6$ symmetry. The calculated energy per 5-7 pair (dislocation core) is the smallest for any known topological defect, suggesting that this defect is likely to form under conditions where mobile dislocations exist (during the preparation of this manuscript, the flower defect was identified independently in CVD-grown graphene on Ni[21,22]).

Apart from the Stone-Wales defect, other rotational grain boundaries have yet to be experimentally observed, but may prove to be very interesting with regard to their influence on electronic transport[16] and mechanical properties.[10]

**Acknowledgements**

We thank S. Adam and E. Mele for useful discussions. PNF acknowledges support from the Semiconductor Research Corporation (NRI-INDEX) and from NSF (DMR-0804908 and DMR-0820382 [MRSEC]).

**Figure Captions**:

Figure 1. Schematic of defects constructed from positive and negative 60° disclinations. (a) A positive 60° disclination with a five-membered ring (blue) in the hexagonal lattice. (b) A negative 60° disclination with a seven-membered ring (pink). (c) A dislocation is produced by joining the complementary disclinations in (a) and (b) with Burgers vector, $\mathbf{b}_{(0,1)}$. (d) Close packing of (0,1) dislocations creates a large angle linear grain boundary with angle $\theta = 21.8°$. (e) A large angle linear grain boundary with $\theta = 32.2°$ produced by combining (0,1) + (1,0) dislocations. (f) Two dislocations from (c) rotated by 60° relative to one another. Topologically closed structures can be formed by rotating a sequence of dislocations, as shown in Fig. 2. Seven and five-fold rings are colored pink and blue, respectively.

Figure 2. Schematic structures of the low-energy rotational grain boundaries identified in this work. (a) Stone-Wales defect with a $C_2$ symmetry axis. (b) Rotational grain boundary with a $C_3$ symmetry axis. (c) Flower defect, with $C_6$ symmetry axis. Seven and five-fold rings are colored pink and blue, respectively.

Figure 3. Examples of low-energy grain boundary loops that do not conserve the number of carbon atoms. (a) 555-777 defect[29] which removes two atoms from the lattice. (b) H_{5,6,7} defect[27] which incorporates two extra atoms into the lattice. Seven and five-fold rings are colored pink and blue, respectively.

Figure 4. Schematic structures of the rotational grain boundary family $C_6(m,n)$ with $C_6$ symmetry. (a) Structure $C_6(1,1)$. (b) Structure $C_6(2,1)$. (c) $C_6(3,1)$. (d) Structure $C_6(2,2)$. The structure in (a) is the flower defect shown in Fig.2 and discussed in the main text. Seven and five-fold rings are colored pink and blue, respectively.

Figure 5. A comparison of experimental and simulated STM topographic images of the $C_6(1,1)$ rotational grain boundary. STM topographs of a 6-fold defect observed in the growth of epitaxial graphene on SiC at (a) -300 mV and (c) 300 mV sample bias. Image size 3.0 nm x 3.5

nm. The color scale intensity (dark to bright) corresponds to topographic height variations of 150 pm and 130 pm for (a) and (c), respectively. Tunneling current setpoint 100 pA. T=4.3 K. (b) Simulated STM image of the $C_6(1,1)$ defect using DFT calculations as detailed in the text. The simulation corresponds to a constant charge density surface of 0.06 eV nm$^{-3}$ obtained by integrating the DOS from $E_D$ to $E_D + 0.3$ eV ($E_F - 0.3$ eV to $E_F$) to simulate the image in (a). The color scale corresponds to a topographic height variation of 140 pm from dark to bright. A lattice model of the $C_6(1,1)$ rotational grain boundary is overlaid onto of the STM simulation.

Figure 6. Electronic properties of the flower defect. (a) A series of STS differential conductance (d$I$/d$V$) spectra measured over a flower defect (inset in (b)). The horizontal axis is the sample bias from -500 mV to 500 mV, and the vertical axis is the distance across the defect from 0 nm to 5 nm. The d$I$/d$V$ intensity is shown in a color scale from 0 nS to 0.4 nS. Superimposed on the STS line map is the topographic profile (white line) taken through the flower defect, showing that spectral peak observed at 200 mV is localized over the flower defect. (b) d$I$/d$V$ spectra averaged over the region of the flower defect (red curve) and in the region outside (blue curve). A minimum in the differential conductance is observed at the Dirac point at -300 mV, and a localized flower state is observed at +200 mV (0.5 eV above $E_D$). The Fermi level is at 0 mV. T=290 K. (Inset) STM image of the flower defect. The d$I$/d$V$ measurements in (a) were obtained along the indicated red line, which is along a pentagon symmetry axis. (c) The calculated projected density of states of the flower rotational grain boundary as a function of spatial position along a pentagon axis (e.g., along a horizontal line through the center of Fig. 2(d)). The defect is centered at the position of 2.5 nm. The data were broadened by Gaussian functions with widths of $\sigma$ =50 meV in energy and $\sigma$ =0.142 nm spatially. (d) The calculated density of states in the vicinity of the Dirac point spatially averaged over the defect (red curve) and for pristine graphene (blue curve). Both spectra were broadened with a Gaussian of width $\sigma$ =50 meV. The dashed yellow lines in (c) and (d) indicate the experimental energy range in (b).

Figure 7. Spatially projected density of states at the various peak energies observed in Fig. 5(d). These images simulate energy-resolved d$I$/d$V$ maps of the flower defect states. The electron density is calculated at 0.33 nm above the graphene surface using DFT results, and then convoluted spatially by a Gaussian of width $\sigma$ = 0.142 nm. Image size is 3 nm x 3 nm. (a) Peak at $E_D$ – 0.4 eV (b;c) Peaks at $E_D$ + 0.2 eV and at $E_D$ + 0.4 eV; (d) peak at $E_D$ + 0.6 eV. The peaks at $E_D$ + 0.2 eV and $E_D$ + 0.4 eV arise from a single localized resonance at $E_D$ + 0.3 eV.

| Figure | Defect | $N_{core}$ | Angle θ | Energy (eV) | Energy per dislocation (5-7 pair) (eV) |
|---|---|---|---|---|---|
| 1d | Large-angle grain boundary | | 21.8° | | 2.2 |
| 1e | Large-angle grain boundary | | 32.2 | | 1.3 |
| 2a | Stone-Wales | 2 | 90° | 4.0 | 2.0 |
| 2b | $C_3$ GB Loop | 13 | ≈40° | 9.3 | 3.1 |
| 2c | $C_6(1,1)$ (Flower defect) | 24 | 30° | 7.0 | 1.2 |
| 4b | $C_6(2,1)$ | 54 | ≈20° | 19.9 | 3.3 |

Table 1: The properties of various linear and rotational grain boundaries (GB) in graphene. Rotation of $N_{core}$ central atoms by the angle θ yields the rotational grain boundaries in the indicated figures. Energies are relative to the pristine graphene lattice, as determined by DFT calculations.

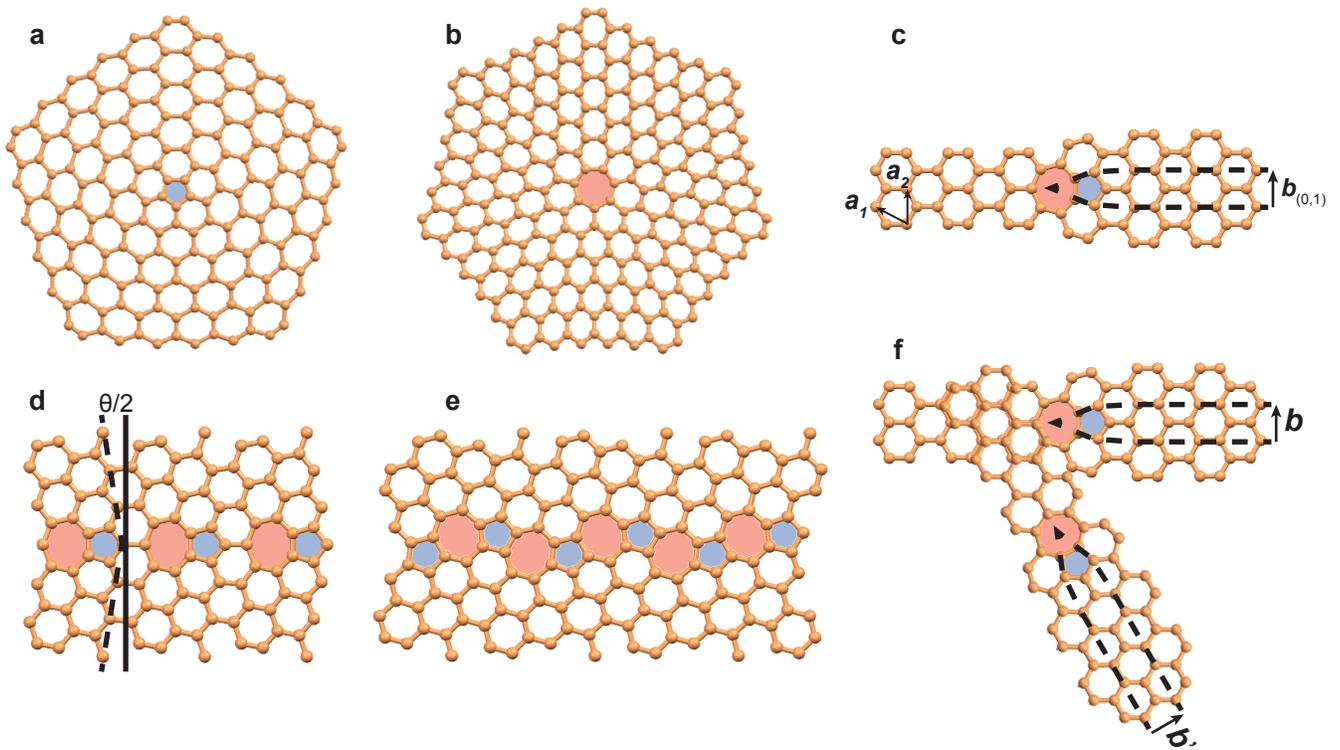

Fig. 1. Schematic of defects constructed from positive and negative 60° disclinations. (a) A positive 60° disclination with a five-membered ring (blue) in the hexagonal lattice. (b) A negative 60° disclination with a seven-membered ring (pink). (c) A dislocation is produced by joining the complimentary disclinations in (a) and (b) with Burgers vector, $\mathbf{b}_{(0,1)}$. (d) Close packing of (0,1) dislocations creates a large angle linear grain boundary with angle $\theta = 21.8°$. (e) A large angle linear grain boundary with $\theta = 32.2°$ produced by combining (0,1) + (1,0) dislocations. (f) Two dislocations from (c) rotated by 60° relative to one another. Topologically closed structures can be formed by rotating a sequence of dislocations, as shown in Fig. 2. Seven and five-fold rings are colored pink and blue, respectively.

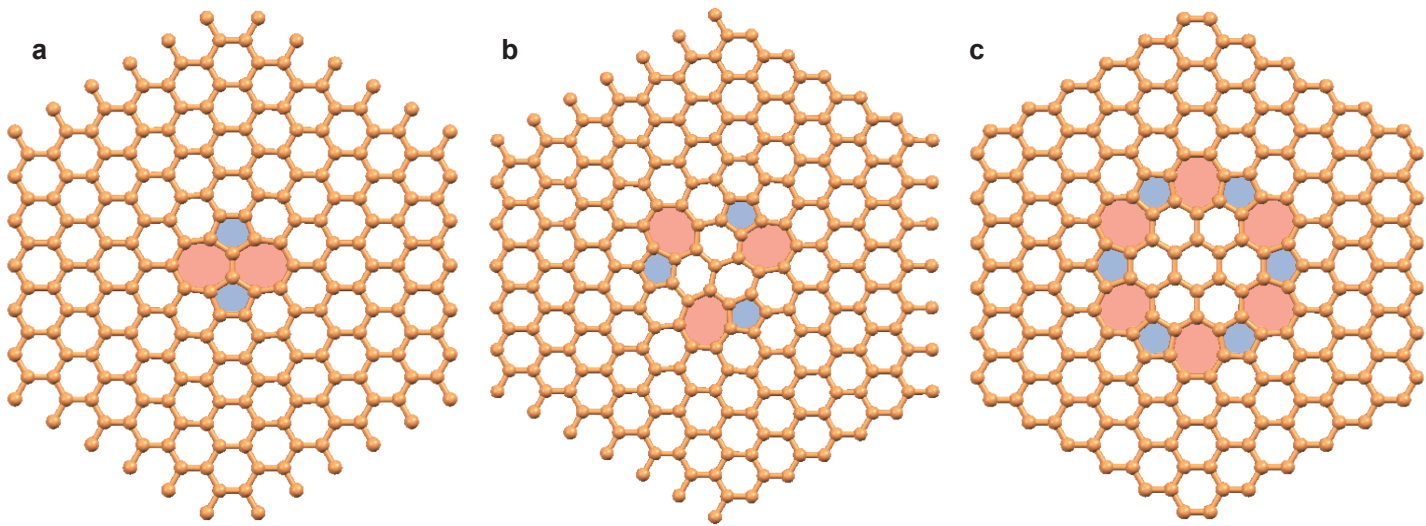

Fig. 2. Schematic structures of the low-energy rotational grain boundaries identified in this work. (a) Stone-Wales defect with a $C_2$ symmetry axis. (b) Rotational grain boundary with a $C_3$ symmetry axis. (c) Flower defect, with $C_6$ symmetry axis. Seven and five-fold rings are colored pink and blue, respectively.

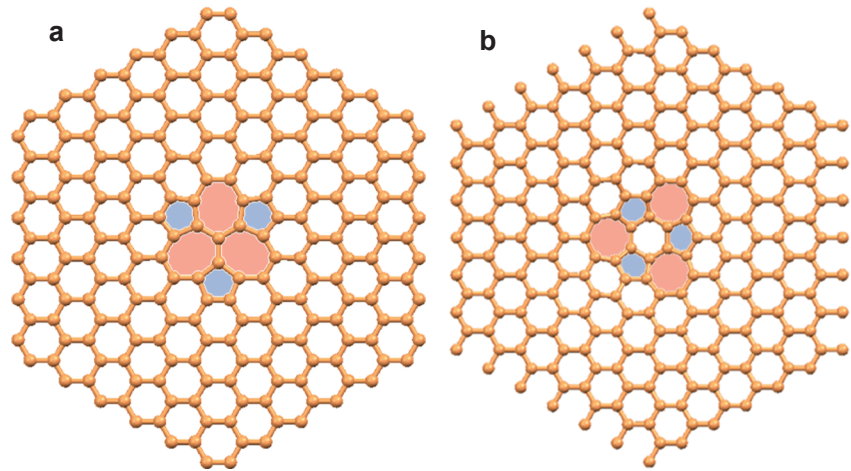

Fig. 3. Examples of low-energy grain boundary loops that do not conserve the number of carbon atoms. (a) 555-777 defect[29] which removes two atoms from the lattice. (b) H_{5,6,7} defect[27] which incorporates two extra atoms into the lattice. Seven and five-fold rings are colored pink and blue, respectively.

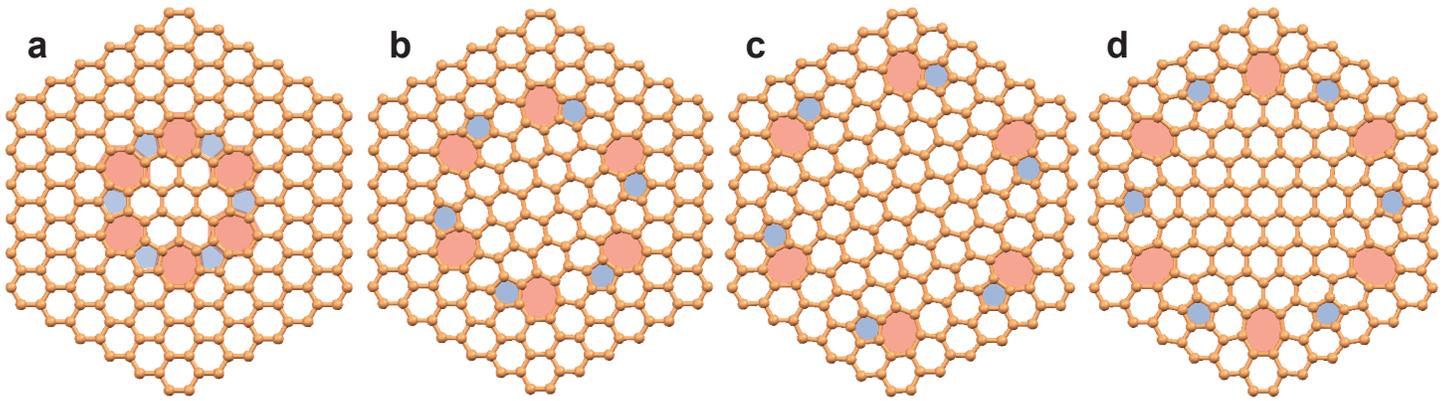

Fig. 4. Schematic structures of the rotational grain boundary family $C_6(m,n)$ with $C_6$ symmetry. (a) Structure $C_6(1,1)$. (b) Structure $C_6(2,1)$. (c) $C_6(3,1)$. (d) Structure $C_6(2,2)$. The structure in (a) is the flower defect shown in Fig. 2 and discussed in the main text. Seven and five-fold rings are colored pink and blue, respectively.

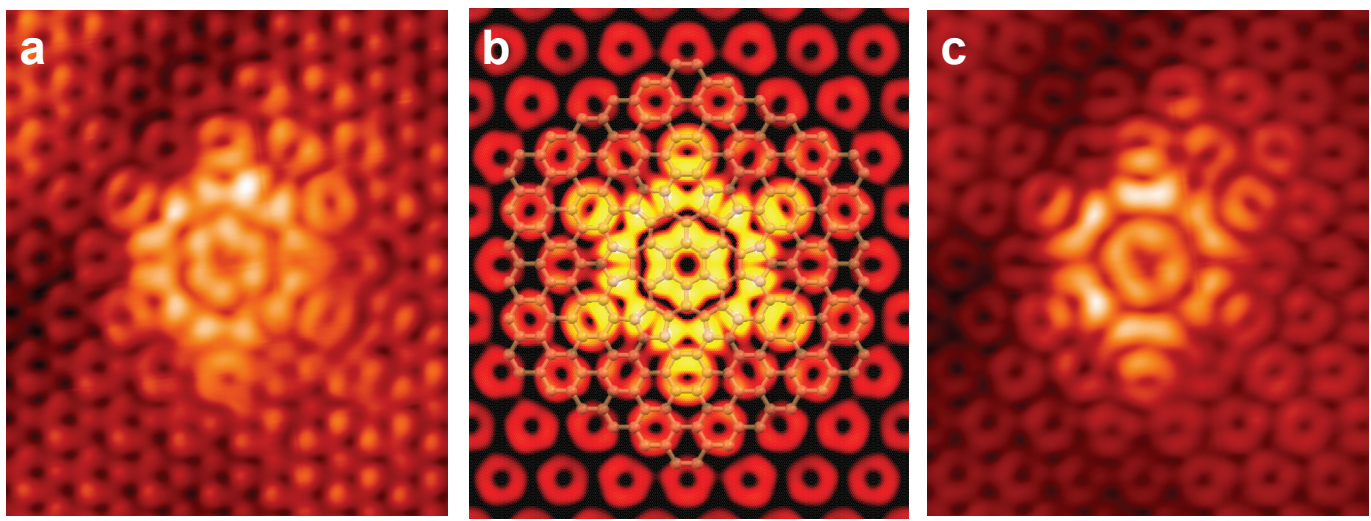

Fig. 5. A comparison of experimental and simulated STM topographic images of the $C_6(1,1)$ rotational grain boundary. STM topographs of a 6-fold defect observed in the growth of epitaxial graphene on SiC at (a) -300 mV and (c) 300 mV sample bias. Image size 3.0 nm x 3.5 nm. The color scale intensity (dark to bright) corresponds to topographic height variations of 150 pm and 130 pm for (a) and (c), respectively. Tunneling current setpoint 100 pA. T=4.3 K. (b) Simulated STM image of the $C_6(1,1)$ defect using DFT calculations as detailed in the text. The simulation corresponds to a constant charge density surface of 0.06 eV nm$^{-3}$ obtained by integrating the DOS from $E_D$ to $E_D + 0.3$ eV ($E_F - 0.3$ eV to $E_F$) to simulate the image in (a). The color scale corresponds to a topographic height variation of 140 pm from dark to bright. A lattice model of the $C_6(1,1)$ rotational grain boundary is overlaid onto of the STM simulation.

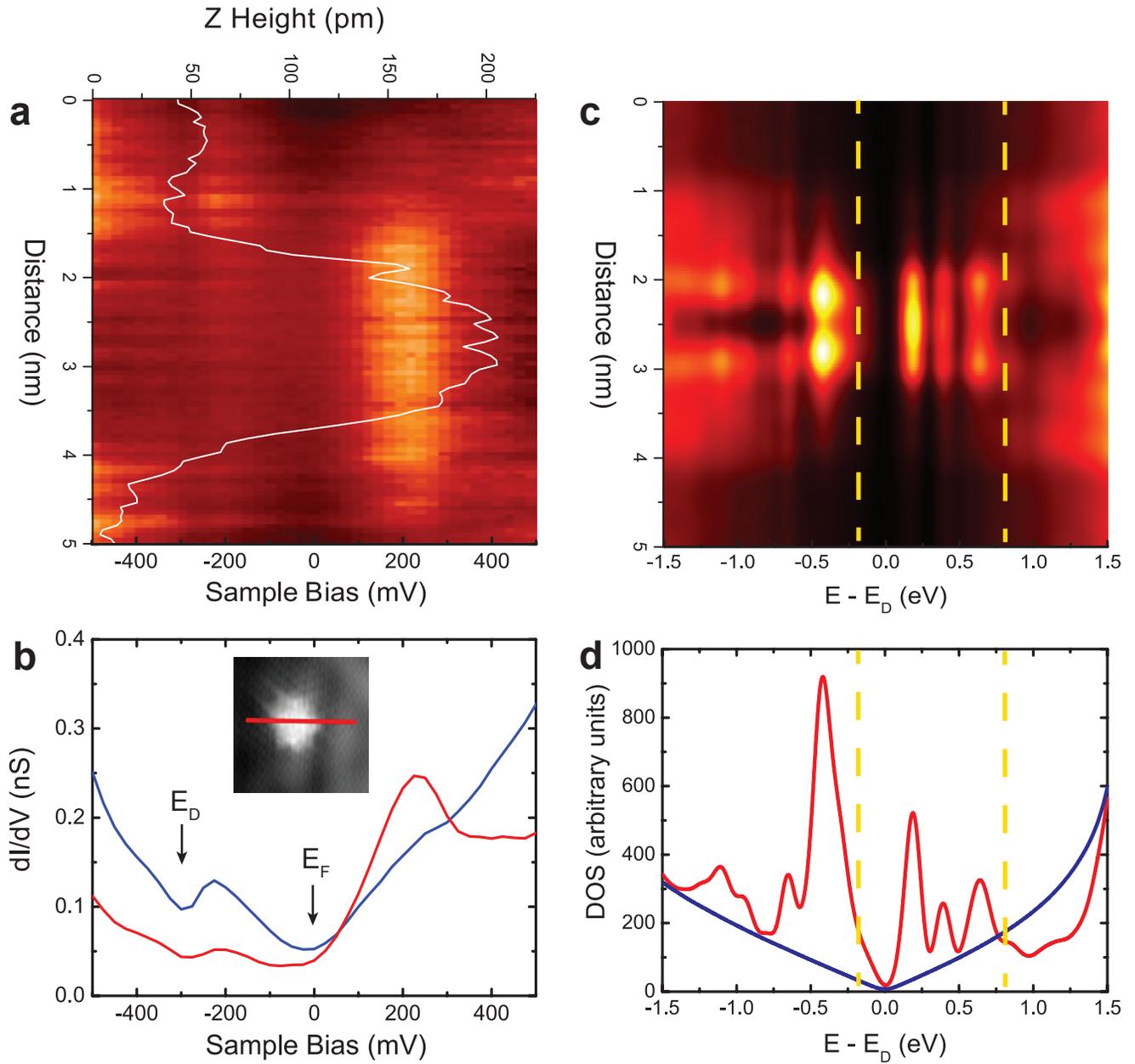

Fig. 6. Electronic properties of the flower defect. (a) A series of STS differential conductance (d$I$/d$V$) spectra measured over a flower defect (inset in (b)). The horizontal axis is the sample bias from -500 mV to 500 mV, and the vertical axis is the distance across the defect from 0 nm to 5 nm. The d$I$/d$V$ intensity is shown in a color scale from 0 nS to 0.4 nS. Superimposed on the STS line map is the topographic profile (white line) taken through the flower defect, showing that spectral peak observed at 200 mV is localized over the flower defect. (b) d$I$/d$V$ spectra averaged over the region of the flower defect (red curve) and in the region outside (blue curve). A minimum in the differential conductance is observed at the Dirac point at -300 mV, and a localized flower state is observed at +200 mV (0.5 eV above $E_D$). The Fermi level is at 0 mV. T=290 K. (Inset) STM image of the flower defect. The d$I$/d$V$ measurements in (a) were obtained along the indicated red line, which is along a pentagon symmetry axis. (c) The calculated projected density of states of the flower rotational grain boundary as a function of spatial position along a pentagon axis (e.g., along a horizontal line through the center of Fig. 2(d)). The defect is centered at the position of 2.5 nm. The data were broadened by Gaussian functions with widths of σ=50 meV in energy and σ=0.142 nm spatially. (d) The calculated density of states in the vicinity of the Dirac point spatially averaged over the defect (red curve) and for pristine graphene (blue curve). Both spectra were broadened with a Gaussian of width σ=50 meV. The dashed yellow lines in (c) and (d) indicate the experimental energy range in (b).

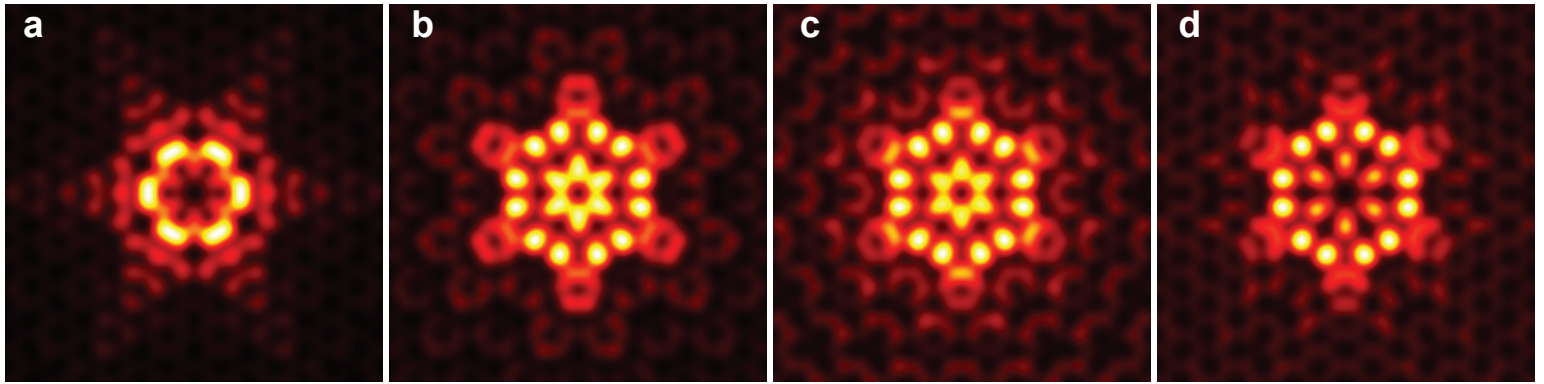

Fig. 7. Spatially projected density of states at the various peak energies observed in Fig. 5(d), which simulate energy resolved STM d$I$/d$V$ mapping of the flower defect states. The electron density is calculated at 0.33 nm above the graphene surface using DFT results, and then convoluted spatially by a Gaussian of width $\sigma$ = 0.142 nm. Image size is 3 nm x 3 nm. (a) Resonance at $E_D$ – 0.4 eV (b;c) Resonance at $E_D$ + 0.3 eV; (d) resonance at $E_D$ + 0.6 eV